\documentclass[10pt,A4paper,conference]{IEEEtran}
\usepackage{epsfig}
\usepackage{amssymb}
\usepackage{verbatim}
\usepackage{delarray}
\usepackage{graphics}
\usepackage{epic}

\newcommand{\inner}[2]{\langle{#1},{#2}\rangle}
\newcommand{\openbox}{\leavevmode
  \hbox to.77778em{%
  \hfil\vrule
  \vbox to.675em{\hrule width.6em\vfil\hrule}%
  \vrule\hfil}}
\newcommand{\qed}{\hspace*{1cm}\hspace*{\fill}\openbox}

\newcommand{\CC}{{\mathcal{C}}}

\newcommand{\F}{{\mathbb{F}}}

\newcommand{\Z}{{\mathbb{Z}}}

\newcommand{\eb}{{\mathbf e}}
\newcommand{\gb}{{\mathbf g}}

\newcommand{\Eb}{{\mathbf E}}

\newcommand{\ie}{{\em i.e., }}
\newcommand{\eg}{{\em e.g., }}

\begin{document}


\title{Simple Rate-1/3 Convolutional and Tail-Biting Quantum Error-Correcting
Codes}


\author{\authorblockN{G. David Forney, Jr.}
\authorblockA{Lab.\ for Inform.\ and Dec.\ Syst.\\
Mass.\ Inst.\ Tech.\\
Cambridge, MA 02139 USA \\
Email: forneyd@comcast.net}
\and
\authorblockN{Saikat Guha}
\authorblockA{Research Lab.\ Electronics\\
Mass.\ Inst.\ Tech.\\
Cambridge, MA 02139 USA \\
Email: saikat@mit.edu}
 }

\maketitle

\begin{abstract}

Simple rate-1/3 single-error-correcting unrestricted and CSS-type quantum
convolutional codes are constructed from classical self-orthogonal
$\F_4$-linear and $\F_2$-linear convolutional codes, respectively.  These
quantum convolutional codes have higher rate than comparable quantum
block codes or previous quantum convolutional codes, and are simple to
decode.  A block single-error-correcting $[9, 3, 3]$
tail-biting code is derived from the unrestricted convolutional code, and
similarly a $[15, 5, 3]$  CSS-type block code from the CSS-type
convolutional code.

\end{abstract}

\section{Introduction}

The field of quantum error-correcting codes (QECCs) has made substantial
progress since the first 9-qubit single-error-correcting code
was proposed by Shor in 1995 \cite{S95}.  More efficient 7-qubit and
5-qubit single-error-correcting codes have been discovered
\cite{NC00}.  A general theory of stabilizer codes has been elucidated
\cite{CRSS97, G98, NC00}.  Within this framework, a theory of
$\F_4$-linear stabilizer codes has been developed \cite{CRSS98}. 
Among these codes are Calderbank-Shor-Steane (CSS) codes
\cite{CS96, S96}, which are based on binary codes.   Using these
structures, a large variety of block QECCs have been proposed.

In classical coding, practical systems have mostly used convolutional
codes rather than block codes, because convolutional codes are usually
superior in terms of their performance-complexity tradeoff.  While this
tradeoff does not seem to have been much of an issue to date for QECCs,
a few attempts have been made to construct quantum convolutional codes
(QCCs).  

Chau \cite{C98, C99} proposed several ``quantum convolutional codes,"
but whether this term is actually appropriate for the Chau codes is
debatable.  Ollivier and Tillich
\cite{OT03, OT04} have given an example of a rate-1/5
single-error-correcting QCC, and have addressed gate-level
implementation issues, but unfortunately their example QCC does not
improve on the comparable 5-qubit block code in
either performance or complexity.  Most recently, Almeida and Palazzo
\cite{AP04} have constructed a rate-1/4 single-error-correcting Shor-type
concatenated QCC;  this code has a higher rate than comparable
block codes, but its encoding and decoding appear to be rather complex.

In this paper, we present via simple examples four new classes of quantum
codes--- namely, $\F_4$-based and CSS-type convolutional and tail-biting codes. 
We claim to exhibit:

\begin{itemize}
\item The first QCCs with clear advantages in both performance and complexity
over comparable block codes;  
\item The first quantum tail-biting codes, with recognition of their complexity
advantages as quantum block codes;
\item The first CSS-type convolutional codes, with recognition of their
complexity advantages over $\F_4$-based codes.
\end{itemize}

Specifically, we present rate-1/3 single-error-correcting 
$\F_4$-based and CSS-type QCCs which have higher rate than any of these
prior single-error-correcting codes, and which are simple to decode. 
Moreover, we derive from these codes simple
tail-biting block codes, which also have rate 1/3, and which can correct
single errors with equally simple decoding algorithms.  In future work, we will
generalize these examples.

In Section II, using the theory of
$\F_4$-linear stabilizer codes developed by Calderbank, Rains, Shor
and Sloane \cite{CRSS98}, we
construct a simple rate-1/3 single-error-correcting quantum
convolutional code from a classical rate-1/3 self-orthogonal
$\F_4$-linear convolutional code.  We
give a simple decoding algorithm for this code that involves only a
9-entry table lookup.  Using tail-biting, we derive a $[9, 3, 3]$ (\ie
9-qubit, rate-1/3, single-error-correcting) block stabilizer code, which
can be decoded by the same simple decoding algorithm.

In Section III, we construct CSS-type codes based on binary codes, which have
certain advantages over unrestricted
$\F_4$-linear codes;  in particular, bit flip and phase flip errors may be
corrected independently.  For example, the Steane 7-qubit code is a CSS-type
code which may be preferred to the 5-qubit single-error-correcting block
code, even though it has lower rate.  Here we present a rate-1/3
single-error-correcting CSS-type quantum convolutional code which is extremely
simple to decode.  We derive from this code a
$[15, 5, 3]$ tail-biting single-error-correcting block code which has the same
rate, and an equally simple decoding algorithm.

\section{Codes based on $\F_4$-linear
codes}

The development of Calderbank, Rains, Shor
and Sloane \cite{CRSS98} leads to
the following proposition:

\smallskip
\textbf{Proposition A}.  Given $n, k$ with $0 \le k \le n$
and $n-k$  even, and given a classical self-orthogonal $(n, (n-k)/2)$
$\F_4$-linear block code $\CC$  over the quaternary field $\F_4$ whose
orthogonal  $(n, (n+k)/2)$ code $\CC^\perp$ under the Hermitian inner
product has minimum  Hamming distance $d$, there exists a quantum
$[n,k,d]$ stabilizer code that encodes $k$ qubits into $n$ qubits and can
correct any pattern of up to
$\lfloor(d-1)/2\rfloor$ qubit errors.

The codes $\CC$ and $\CC^\perp$ are the quaternary label codes $L(S)$
and $L(N(S))$ of the stabilizer group $S$ and the normalizer group
$N(S)$, respectively, where the quaternary labels of the four Pauli
matrices $\{I, X, Y, Z\}$ are respectively the elements $\{0, \omega, 1,
\overline{\omega}\}$ of the quaternary field $\F_4$.

As with a classical code, decoding of a stabilizer code involves
measuring a set of $(n-k)/2$ $\F_4$-syndromes $S_j =
\inner{L(\Eb)}{\gb_j} \in \F_4$, where $\{\gb_j\}$
is a set of $(n-k)/2$ generators of $\CC$, and
$\inner{L(\Eb)}{\gb_j}$ denotes the Hermitian inner product of $\gb_j$
with a quaternary error label sequence $L(\Eb)$.  The syndromes
identify the error label sequence $L(\Eb)$ as belonging to one of
$4^{(n-k)/2}$ cosets of the orthogonal code
$\CC^\perp$.  

The decoder then determines the error label
sequence of minimum Hamming weight in that coset.  If $n-k$ is not too
large, then this can be done by a table lookup in a table with
$4^{(n-k)/2}$ entries.  (The question of decoding  complexity for large
codes seems hardly to have been addressed previously in the QECC
literature, with the notable exception of
\cite{MMM04}.)

\smallskip
\noindent
\textbf{Example A} (Five-qubit ``quantum Hamming code").
There exists a $(5, 2)$ self-orthogonal linear block code $\CC$ over
$\F_4$, generated by $\gb_1 = (0, \overline{\omega}, \omega, \omega,
\overline{\omega})$ and $\gb_2 = (\overline{\omega}, 0,
\overline{\omega}, \omega, \omega)$,
whose orthogonal code $\CC^\perp$ is a $(5, 3, 3)$ linear Hamming code
over $\F_4$. There therefore exists a quantum $[5,1,3]$ code;  \ie
a code that encodes 1 qubit into 5 qubits, and corrects any single error.
Because the 15 possible single-error label sequences  $L(\Eb)$ map
one-to-one to the 15 nonzero cosets of $\CC^\perp$, this is a ``quantum
Hamming code."  Decoding may be done by a table lookup in a table with 16
entries.
\qed

\subsection{A simple rate-1/3, single-error-correcting QCC}

We now construct a rate-1/3
convolutional stabilizer code with minimum Hamming distance $d = 3$ using
Proposition A; \ie we find a classical self-orthogonal
rate-1/3 $\F_4$-linear convolutional code $\CC$ whose
orthogonal  code $\CC^\perp$ under the Hermitian inner product has
minimum distance 3.

Consider the classical rate-1/3 $\F_4$-linear shift-invariant
convolutional code $\CC$
with generators:
$$
\begin{array}{c}
\ldots \\
\begin{array}{c|ccc|ccc|ccc|c}
\ldots & 1 & 1 & 1 & 1 & \omega & \overline{\omega} & 0 & 0 & 0 & \ldots
\\
\ldots & 0 & 0 & 0 & 1 & 1 & 1 & 1 & \omega & \overline{\omega} & \ldots
\end{array} \\
\ldots
\end{array}
$$
In other words, for every block of $n=3$ qubits, there is one
generator, so the classical rate is $1/3$.  The generators are of the
``sliding block" type;  that is, every generator is a shift by an
integral number of blocks of a single basic generator $(\ldots, 000, 111,
1\omega\overline{\omega}, 000, \ldots)$.\footnote{The stabilizer group
$S$ is actually generated by sequences of Pauli matrices that correspond
to multiples by $\omega$ and $\overline{\omega}$ of the above
generators;  \ie the generators of $S$ are the shifts by an integral
number of blocks of the two basic generators $(\ldots, III, XXX, XZY,
III, \ldots)$ and  $(\ldots, III, ZZZ, ZYX, III, \ldots)$.}  In $D$-transform
notation, the code generators are $D^k(1 + D, 1 + \omega D, 1 +
\overline{\omega}D), k \in \Z$.

In principle, $\CC$ has an
infinite number of generators covering an infinite number of blocks.
Later we will discuss methods for making such a code into a finite
block code.  However, the code constraints are localized;  the code
symbols in any block are a function only of the ``current" and
``previous" generators.  Such a convolutional code is said to have a
``memory" or ``constraint length" of one block ($\nu = 1$).

All generators are orthogonal under the
Hermitian inner product, so $\CC$ is self-orthogonal.  We will take
$\CC$ as the quaternary label code $L(S)$ of a convolutional stabilizer
code.

The rate of this convolutional stabilizer code in quantum terms is also
1/3;  \ie the code encodes one qubit stream into a second stream at a
rate of three qubits per original qubit.

The orthogonal code $\CC^\perp$ under the Hermitian inner
product is a rate-2/3 $\F_4$-linear shift-invariant convolutional code
whose generators are as follows (in $D$-transform notation, multiples of
$(\overline{\omega},
\omega, 1)$ and  $(1 + D, 1 + \omega D, 1 + \overline{\omega}D)$ by $D^k$):
$$
\begin{array}{c}
\ldots \\
\begin{array}{c|ccc|ccc|ccc|c}
\ldots & \overline{\omega} & \omega & 1 & 0 & 0 & 0 & 0 & 0 & 0 &
\ldots \\
\ldots & 1 & 1 & 1 & 1 & \omega & \overline{\omega} & 0 & 0 &
0 & \ldots \\
\hline
\ldots & 0 & 0 & 0 & \overline{\omega} &
\omega & 1 & 0 & 0 & 0 & \ldots \\
\ldots & 0 & 0 & 0 & 1 & 1 & 1 & 1 & \omega & \overline{\omega} & \ldots
\\
\hline
\end{array} \\
\ldots
\end{array}
$$

The minimum Hamming distance of $\CC^\perp$ is 3, and the only
codewords of weight 3 are single-block codewords.  This is easily seen
because
$(\overline{\omega} \omega 1)$ and $(1 1 1)$ generate a $(3,2,2)$
$\F_4$-linear block code, so every codeword accumulates a Hamming weight
of at least 2 in its first block;  similarly, every codeword accumulates
a Hamming weight of at least 2 in its last block.  The only single-block
codewords are multiples of $(\overline{\omega} \omega 1)$, which have
Hamming weight 3.  
The convolutional stabilizer
code defined by $\CC$ thus has minimum Hamming distance 3, so it is a
single-error-correcting code.

\subsection{Decoding algorithms}

For decoding, we first measure each
generator $\gb_j$ of $\CC$ to obtain a sequence of
$\F_4$-syndromes $S_j = \inner{L(\Eb)}{\gb_j} \in \F_4$, where $L(\Eb)$
denotes the quaternary error label sequence $L(\Eb)$, at a rate of one
$\F_4$-syndrome for each block.  This determines a coset of
the orthogonal convolutional code $\CC^\perp$.  We then need to find the
minimum-weight coset leader in that coset.

For any convolutional code, a standard way of finding coset leaders  is by a
Viterbi algorithm (VA) search.  It can easily be seen that
$\CC^\perp$ has a trellis diagram with 4 states at each
block boundary, and with 64 transitions between trellis states during
each block.  A VA search through such a trellis is not difficult,
but requires of the order of 64 computations per block.

If our objective is merely correction of single errors, however, then we
can use a much simpler algorithm, as follows.  As long as all syndromes
are zero, we assume that no errors have occurred.  Then, if a nonzero
syndrome $S_j$ occurs, we assume that a single error has occurred in one
of the three qubits in block $j$;  the error is characterized by a label
3-tuple $\eb_j = L(\Eb_j)$.  The nine possible weight-1 error 3-tuples 
$\eb_j$ lead to the following syndromes $(S_j,S_{j+1})$:

$$
\begin{array}{c|c}
\eb_j & (S_j, S_{j+1}) \\
\hline
1 0 0  & (1, 1) \\
\omega 0 0 & (\omega, \omega) \\
\overline{\omega} 0 0 & (\overline{\omega}, \overline{\omega}) \\
0 1 0  & (\overline{\omega}, 1) \\
0 \omega 0 & (1, \omega) \\
0 \overline{\omega} 0 & (\omega, \overline{\omega}) \\
0 0 1  & (\omega, 1) \\
0 0 \omega & (\overline{\omega}, \omega) \\
0 0 \overline{\omega} & (1, \overline{\omega}) 
\end{array}
$$
Since these nine syndrome pairs $(S_j,S_{j+1})$ are distinct, we can map
$(S_j,S_{j+1})$ to the corresponding single-error label 3-tuple $\eb_j$
using a simple 9-entry table lookup, and then correct the error as
indicated.  (If $S_{j+1} = 0$, \ie if $S_j$ is an isolated nonzero syndrome, then
we have detected a weight-2 error.)

We see that this simple algorithm can correct any single-error
pattern $\Eb_j$, provided that there is no second error during blocks
$j$ and $j+1$.   The decoder synchronizes itself properly whenever a zero
syndrome occurs, and subsequently can correct one error in every
second block, provided that every errored block is followed by an
error-free block.

\subsection{Terminated and tail-biting block codes}

A standard method for reducing a convolutional code to a block code
without loss of minimum distance is to terminate it;  \ie to take as the
block code the set of all convolutional code sequences that are nonzero
only during a given interval of $N$ blocks.  The resulting code is a
linear block code which is a subcode of the convolutional code,
and thus has at least the same minimum distance.

For example, if $\CC^\perp$ is terminated to an
interval of $N$ blocks, then it becomes an $\F_4$-linear block code with
parameters $(3N,2N-1,3)$, because there are $2N-1$ generators that are
nonzero only in the defined $N$-block interval.  For instance, if
$N=3$, then we obtain a classical linear $(9, 5, 3)$ block code, which
yields a quantum $[9,1,3]$ stabilizer code.  As $N \to \infty$, the
classical rate approaches 2/3, and the corresponding quantum rate
approaches $1/3$.

Another, better idea for creating a linear block code from  $\CC^\perp$
is to use tail-biting, which preserves rate but possibly not minimum
distance.  For tail-biting, we take the set of all generators that
``start" during a given interval of $N$ blocks, and wrap around any
blocks that do not fit within the given interval back to the beginning
in cyclic ``end-around" fashion.

For our orthogonal code $\CC^\perp$, it turns out that
there is no loss of minimum distance whenever $N \ge 3$.  In particular,
the following set of tail-biting generators generate a $(9,6,3)$
$\F_4$-linear block code, which is the normalizer label code of a quantum
$[9, 3, 3]$ stabilizer code:
$$
\begin{array}{ccc|ccc|ccc}
\overline{\omega} & \omega & 1 & 0 & 0 & 0 & 0 & 0 & 0 \\
1 & 1 & 1 & 1 & \omega & \overline{\omega} & 0 & 0 & 0 \\
\hline
0 & 0 & 0 & \overline{\omega} & \omega & 1 & 0 & 0 & 0 \\
0 & 0 & 0 & 1 & 1 & 1 & 1 & \omega & \overline{\omega} \\
\hline
0 & 0 & 0 & 0 & 0 & 0 & \overline{\omega} & \omega & 1 \\
1 & \omega & \overline{\omega} & 0 & 0 & 0 & 1 & 1 & 1 \\
\hline
\end{array}
$$
The second, fourth and sixth generators generate the dual $(9, 3)$
tail-biting stabilizer label code.

To decode this code, we can use the same decoding algorithm as before,
but now on a ``circular" time axis.  Specifically, if only a single error
occurs, then one of the three resulting $\F_4$-syndromes will be zero,
and the other two nonzero.  The zero syndrome tells which block the
error is in;  the remaining two nonzero syndromes determine the error
pattern according to the 9-entry table given earlier.  Thus again we
need only a 9-entry table lookup.

\subsection{Error probability}

We now briefly consider decoding error probability.  We assume that the
probability of an error in any qubit is $p$, independent of errors in
other qubits.  Our estimates do not depend on the relative probabilities
of $X, Y$ or $Z$ errors. 

For the 5-qubit block code of Example A, a decoding error may occur if
there are 2 errors in any block, so the error probability is of
the order of ${5 \choose 2} p^2 = 10p^2$ per block, or per encoded qubit.

For the rate-1/3 convolutional code, for each 3-qubit block, a decoding
error may occur if there are 2 errors in that block, or 1 in that block
and 1 in the subsequent block.  The error probability is therefore of
the order of $(3 + 3^2) p^2 = 12 p^2$ per 3-qubit block, or per encoded
qubit.

Finally, for the $[9, 3, 3]$ tail-biting block code, a decoding error
may occur if there are 2 errors in a block of 9 qubits, so the error
probability is of the  order of ${9 \choose 2} p^2 = 36p^2$ per block,
or $12p^2$ per encoded qubit.

We conclude that the decoding error probability is very nearly the same
for any of these codes.

\subsection{Discussion}

Our quantum convolutional code has rate 1/3, which is greater than that
of any previous simple single-error-correcting quantum code, block or
convolutional.  Our decoding algorithm involves only a 9-entry table
lookup, which is at least as simple as that of any previous quantum code.

Our convolutional code rate and error-correction capability are
comparable to those of a $[6, 2, 3]$ block stabilizer code.  However, by
the ``quantum Hamming bound," there exists no $[6, 2, 3]$ block
stabilizer code.

Our tail-biting code is a $[9,3,3]$ block stabilizer code.  A code with
the same parameters may be obtained by shortening a $[21, 15, 3]$
quantum Hamming code.  However, such a shortened code would not have
such a simple structure as our tail-biting code, nor such a simple
decoding algorithm.

\section{CSS-type codes}

The binary field $\F_2$ is a subfield of the quaternary field $\F_4$. 
The $(n-k)/2$ generators of a classical
self-orthogonal $(n, (n-k)/2)$ $\F_2$-linear code may therefore be taken
as the generators of a self-orthogonal $(n, (n-k)/2)$ $\F_4$-linear code
as in Proposition A.  The resulting quantum stabilizer code is then of
the type proposed by Calderbank and Shor \cite{CS96} and Steane
\cite{S96}, which we call a \emph{CSS-type code}.

\smallskip
\textbf{Proposition B}.  Given $n, k$ with $0 \le k \le n$
and $n-k$  even, and given a classical self-orthogonal $(n, (n-k)/2)$
$\F_2$-linear block code $\CC$  over the binary field $\F_2$ whose
orthogonal  $(n, (n+k)/2)$ code $\CC^\perp$ has minimum  Hamming
distance $d$, there exists an
$[n,k,d]$ CSS-type code.

For CSS-type codes, we may think of the four Pauli matrices $\{I, X, Y,
Z\}$ as having two-bit labels $\{00, 10, 11, 01\}$, respectively.  The
first bit is called the bit flip bit, and the second the phase flip
bit.  Thus an $X$ error is a bit flip error, a $Z$ error is a phase flip
error, and a $Y$ error is a combined bit and phase flip error.

CSS-type codes have the advantage that these two types of error bits are
protected by two independent binary codes, which may be independently
decoded.  On the other hand, the parameters $[n,k,d]$ of
CSS-type codes are not generally as good as those of unrestricted
codes, because the parameters of binary codes are not generally as good
as those of quaternary codes.

Decoding of a CSS-type code involves
measuring a set of $(n-k)/2$ pairs of $\F_2$-syndromes $(s_{1,j},
s_{2,j}) = (\inner{\ell_1(\Eb)}{\gb_j}, \inner{\ell_2(\Eb)}{\gb_j}) \in
(\F_2)^2$, where $\{\gb_j\}$ is a set of $(n-k)/2$
generators of the binary code $\CC$, and
$\inner{\ell_1(\Eb)}{\gb_j}$ and $\inner{\ell_2(\Eb)}{\gb_j}$ denote
the binary inner products of
$\gb_j$ with the two binary label sequences $\ell_1(\Eb), \ell_2(\Eb)$,
which respectively denote sequences of bit flip and phase flip errors. 
These syndromes identify each of the two error label sequences
$\ell_1(\Eb),
\ell_2(\Eb)$ as belonging to one of $2^{(n-k)/2}$ cosets of the
orthogonal code
$\CC^\perp$.  

Two identical decoders may operate independently on each of these two
syndrome sequences to determine the two error bit sequences of
minimum Hamming weight in these respective cosets.  If
$n-k$ is not too large, then this can be done by two table lookups in a
table with $2^{(n-k)/2}$ entries.  Thus the decoding complexity is
roughly twice the square root of the decoding complexity for a
comparable quaternary code.

\smallskip
\noindent
\textbf{Example B} (Seven-qubit Steane code \cite{S96}).
There exists a $(7, 3)$ self-orthogonal linear block code $\CC$ over
$\F_2$, 
whose orthogonal code $\CC^\perp$ is a $(7, 4, 3)$ linear Hamming code
over $\F_2$. Thus there exists a $[7,1,3]$ CSS-type code.
Decoding may be done by two table lookups in an 8-entry table.
\qed

\subsection{A simple rate-1/3, single-error-correcting CSS-type QCC}

  In this section we will construct a rate-1/3 CSS-type
convolutional stabilizer code with minimum Hamming distance $d = 3$ using
Proposition B.  That is, we will find a binary self-orthogonal
rate-1/3 linear convolutional code whose
orthogonal code has minimum distance 3.

Consider the binary rate-1/3 
convolutional code $\CC$ whose generators are as follows:
$$
\begin{array}{c}
\ldots \\
\begin{array}{c|ccc|ccc|ccc|ccc|c}
\ldots & 1 & 1 & 1 & 1 & 0 & 0 & 1 & 1 & 0 & 0 & 0 & 0 & \ldots
\\
\ldots & 0 & 0 & 0 & 1 & 1 & 1 & 1 & 0 & 0 & 1 & 1 & 0 & \ldots
\end{array} \\
\ldots
\end{array}
$$
(or $D^k(1 + D + D^2, 1 + D^2, 1), k
\in \Z$, in $D$-transform notation). In other words, the classical rate is $1/3$,
and every generator is a shift by an integral number of blocks of a single basic
generator
$(\ldots, 000, 111, 100, 110, 000, \ldots)$.  Thus $\CC$ has a ``memory" of two
blocks (constraint length $\nu = 2$).\footnote{
The stabilizer group $S$ is actually generated by sequences of Pauli
matrices that correspond to multiples of
the above generators by $\omega$ and $\overline{\omega}$.  Thus the
generators of $S$ are the shifts by an integral number of blocks of two
basic generators,
$(\ldots, III, XXX, XII, XXI, III, \ldots)$ and  $(\ldots, III, ZZZ,
ZII, ZZI, III, \ldots)$.  Note that these stabilizers affect only bit
flip bits and phase flip bits, respectively.}

Each generator is orthogonal to all generators under the
usual binary inner product, so the code is self-orthogonal.  The
generators of the orthogonal rate-2/3 binary convolutional code
$\CC^\perp$ are the shifts of two basic generators (in $D$-transform notation, 
multiples  by $D^k$ of
$(1, 1+D, D)$ and  $(D,D,1)$):
$$
\begin{array}{c}
\ldots \\
\begin{array}{c|ccc|ccc|ccc|c}
\ldots & 1 & 1 & 0 & 0 & 1 & 1 & 0 & 0 & 0 & \ldots \\
\ldots & 0 & 0 & 1 & 1 & 1 & 0 & 0 & 0 & 0 & \ldots \\
\hline
\ldots & 0 & 0 & 0 & 1 & 1 & 0 & 0 & 1 & 1 & \ldots \\
\ldots & 0 & 0 & 0 & 0 & 0 & 1 & 1 & 1 & 0 & \ldots \\
\hline
\end{array} \\
\ldots
\end{array}
$$
It is easily verified that the minimum distance of $\CC^\perp$ is $d=3$.

Following Proposition B, we thus obtain from the binary self-orthogonal
convolutional code $\CC$ a CSS-type QCC of
quantum rate $1/3$ and minimum distance $d = 3$.

\subsection{Decoding algorithms}

  We will discuss only decoding of
bit flip errors;  phase flip errors are corrected independently and
identically.

For decoding of bit flip errors, we measure each generator
$\gb_j$ of $\CC$ to obtain a sequence of  binary syndromes $s_{j} =
\inner{\ell_1(\Eb)}{\gb_j}$,  the binary inner products of
the generators
$\gb_j$ with the bit flip error label sequence $\ell_1(\Eb)$, at
a rate of one binary syndrome for each block.

Again, rather than VA decoding the 4-state trellis of the rate-2/3 code 
$\CC^\perp$, we use a simple single-error-correction algorithm, as
follows.  As long as all syndromes
are zero, we assume that no errors have occurred.  Then, if a nonzero
syndrome $s_{j}$ occurs, we assume that a single error has occurred in
one of the three bit flip bits in block $j$;  the error is characterized
by a label 3-tuple $\eb_j = \ell_1(\Eb_j)$.  The three possible weight-1
error 3-tuples 
$\eb_j$ lead to the following bit flip syndromes:
$$
\begin{array}{c|c}
\eb_j & (s_j, s_{j+1}, s_{j+2}) \\
\hline
1 0 0  & (1, 1, 1) \\
0 1 0  & (1, 0 ,1) \\
0 0 1  & (1, 0, 0) 
\end{array}
$$
Since the three syndrome pairs $(s_{j+1}, s_{j+2})$ are distinct,
we can map $(s_{j+1}, s_{j+2})$ to the corresponding single-error
pattern $\eb_j$ using a simple 3-entry table lookup, and then correct
it. 

We see that this simple algorithm can correct any single-error
pattern $\eb_j$, provided that there is no second error during blocks
$j$ through $j+2$.   The decoder synchronizes itself properly whenever a
zero syndrome occurs, and subsequently can correct one error in every
third block, provided that every errored block is followed by two
error-free blocks.

\subsection{Terminated and tail-biting block codes}

For our normalizer code $\CC^\perp$, it turns out that a tail-biting
termination after $N$ blocks results in no loss of minimum
distance whenever
$N \ge 5$.  In particular, the following set of tail-biting generators
generate a
$(15,10,3)$ binary linear block code, which is the normalizer
label code of a quantum
$[15,5, 3]$ CSS-type code:
$$
\begin{array}{ccc|ccc|ccc|ccc|ccc}
1 & 1 & 0 & 0 & 1 & 1 & 0 & 0 & 0 & 0 & 0 & 0 & 0 & 0 & 0 \\
0 & 0 & 1 & 1 & 1 & 0 & 0 & 0 & 0 & 0 & 0 & 0 & 0 & 0 & 0 \\
\hline
0 & 0 & 0 & 1 & 1 & 0 & 0 & 1 & 1 & 0 & 0 & 0 & 0 & 0 & 0 \\
0 & 0 & 0 & 0 & 0 & 1 & 1 & 1 & 0 & 0 & 0 & 0 & 0 & 0 & 0 \\
\hline
0 & 0 & 0 & 0 & 0 & 0 & 1 & 1 & 0 & 0 & 1 & 1 & 0 & 0 & 0 \\
0 & 0 & 0 & 0 & 0 & 0 & 0 & 0 & 1 & 1 & 1 & 0 & 0 & 0 & 0 \\
\hline
0 & 0 & 0 & 0 & 0 & 0 & 0 & 0 & 0 & 1 & 1 & 0 & 0 & 1 & 1 \\
0 & 0 & 0 & 0 & 0 & 0 & 0 & 0 & 0 & 0 & 0 & 1 & 1 & 1 & 0 \\
\hline
0 & 1 & 1 & 0 & 0 & 0 & 0 & 0 & 0 & 0 & 0 & 0 & 1 & 1 & 0 \\
1 & 1 & 0 & 0 & 0 & 0 & 0 & 0 & 0 & 0 & 0 & 0 & 0 & 0 & 1 \\
\hline
\end{array}
$$

To decode this code, we can use the same simple decoding algorithm as for
the corresponding convolutional code, but now on a ``circular" time
axis.  If only a single error occurs, then the first syndrome 1 after
two zeroes (on a circular time axis) identifies the 3-tuple block of the
error, and the next two bits determine its position within the block,
according to the 3-entry table above.

\subsection{Error probability}

Again, we estimate the decoding error probabilities for these
codes when qubit errors are independent and have probability $p$.  We do
not take into account that, because of the independence of the two
decoders, there are some weight-2 error patterns that can be corrected
(\eg $X$ and $Z$);  this would yield a minor improvement in our
estimates.

For the 7-qubit block code of Example B, a decoding error may occur if
there are 2 errors in any block, so the error probability is of
the order of ${7 \choose 2} p^2 = 21p^2$ per block, or per encoded qubit.

For the rate-1/3 convolutional code, for each 3-qubit block, a decoding
error may occur if there are 2 errors in that block, or 1 in that block
and 1 in the two subsequent blocks.  The error probability is therefore
of the order of $(3 + 3\cdot 6) p^2 = 21 p^2$ per 3-qubit block, or per
encoded qubit.

Finally, for the $[15, 5, 3]$ tail-biting block code, a decoding error
may occur if there are 2 errors in a block of 15 qubits, so the error
probability is of the  order of ${15 \choose 2} p^2 = 105p^2$ per block,
or $21p^2$ per encoded qubit.

Again, we conclude that the decoding error probability is very nearly the
same for any of these codes, and is about twice that of the
codes of Section II.

\subsection{Discussion}

Our CSS-type quantum convolutional code has rate 1/3, which is greater
than that of any previous simple CSS-type single-error-correcting quantum
code, block or convolutional.  Our decoder only
requires using a 3-entry table lookup twice.  It is arguably simpler
than that of Section II.

Our convolutional code rate and error-correction capability are
comparable to those of a
$[9, 3, 3]$ CSS-type block code.  However, no $[9, 3, 3]$ CSS-type
block code exists, since there exists no $(9,6,3)$ binary linear block
code, by the classical Hamming bound.

Our tail-biting code is a $[15,5,3]$ CSS-type block code.  A code
with the same parameters may be obtained by shortening a $[31, 21, 3]$
CSS-type block code.  However, such a shortened code would not have such
a simple structure as our tail-biting code, nor such a simple decoding
algorithm.

\section{Future work}

Using the same code construction principles, we have found rate-1/3 $\F_4$-based
and CSS-type codes with up to $1024$ states and
minimum distances up to 8.  We expect to present further examples of
such codes at the ISIT.

\section*{Acknowledgments}

We wish to acknowledge helpful comments by Robert Calderbank and
David MacKay.
Saikat Guha acknowledges the support of Prof.\ Jeffrey H. Shapiro and
the U.S. Army Research Office (DoD MURI Grant No. DAAD-19-00-1-0177).

\end{document}